# There's Plenty of Room Right Here:
## Biological Systems as Evolved, Overloaded, Multi-scale Machines

Joshua Bongard[1,3†] and Michael Levin[2,3†]*

[1] Department of Computer Science, University of Vermont, Burlington, VT USA
[2] Allen Discovery Center at Tufts University, Medford MA, USA.
[3] Institute for Computer Designed Organisms

†Both authors contributed equally to this work.

* Author for correspondence
    200 Boston Ave., Suite 4600
    Medford, MA 02155
    Email: michael.levin@tufts.edu
    Tel. (617) 627-6161

**Abstract**

The applicability of computational models to the biological world is an active topic of debate. We argue that a useful path forward results from abandoning hard boundaries between categories and adopting an observer-dependent, pragmatic view. Such a view dissolves the contingent dichotomies driven by human cognitive biases (e.g., tendency to oversimplify) and prior technological limitations in favor of a more continuous, gradualist view necessitated by the study of evolution, developmental biology, and intelligent machines. Form and function are tightly entwined in nature, and in some cases, in robotics as well. Thus, efforts to re-shape living systems for biomedical or bioengineering purposes require prediction and control of their function at multiple scales. This is challenging for many reasons, one of which is that living systems perform multiple functions in the same place at the same time. We refer to this as "polycomputing" – the ability of the same substrate to simultaneously compute different things. This ability is an important way in which living things are a kind of computer, but not the familiar, linear, deterministic kind; rather, living things are computers in the broad sense of computational materials as reported in the rapidly-growing physical computing literature. We argue that an observer-centered framework for the computations performed by evolved and designed systems will improve the understanding of meso-scale events, as it has already done at quantum and relativistic scales. To develop our understanding of how life performs polycomputing, and how it can be convinced to alter one or more of those functions, we can create technologies that polycompute, and learn how to alter their functions first. Here, we review examples of biological and technological polycomputing, and develop the idea that overloading of different functions on the same hardware is an important design principle that helps understand and build both evolved and designed systems. Learning to hack existing polycomputing substrates, as well as evolve and design new ones, will have massive impacts on regenerative medicine, robotics, and computer engineering.

# 1. Introduction

In Feynman's famous lecture titled "There's Plenty of Room at the Bottom" [1], he argued that vast technological progress could be achieved by learning to manipulate matter and energy at ever smaller scales. Such potential presumably could be exploited by natural selection as well. How does biology expand the adaptive function of an existing system? It can't go down, since there's already something there, exhibiting functional competencies, at every level [2]. Instead, it squeezes more action out of each level by overloading mechanisms with multiple functions – which we term polycomputing. We argue that the most effective lens on a wide range of natural and engineered systems must enable a multiple-observers view where the same set of events can be interpreted as different computations (Figure 1 illustrates how artists have recognized this feature).

Herein we review remarkable examples of biological polycomputing, such as spider webs that serve as auditory sensors and prey capture devices [3], and holographic memory storage in the brain [4,5]. We will also review emerging examples in computer and materials engineering [6]. We provisionally define polycomputing as the ability of a material to provide the results of more than one computation in the same place at the same time. To distinguish this from complex materials that necessarily produce complex results in the same place at the same time (like the multiple peaks in the frequency spectrum of a vibrating material), polycomputing must be embodied in a material that has been evolved, or can be designed to produce particular results – such as the results of particular mathematical transformations like digital logic – and must be readable by other parts of the material or other devices. That is, the computation, to be a computation, must be useful. These advances in understanding polycomputing in biology suggest ways to improve synthetic polycomputing systems, the latter of which in turn shed light on the nature of computation, evolution, and control. Biological systems that polycompute also contributes to an ongoing conceptual debate in interdisciplinary science --- the applicability of computer frameworks and metaphors to living systems [7] --- in three ways. First: if polycomputing changes our understanding of what computation is, that might change whether we consider a living system to be a computer (Sect. 1.1). Second: a living system (or inorganic material) may be considered to be polycomputing depending on one's point of view, suggesting observer dependence is unavoidable when considering whether or what a living or engineered computes (Sect 1.2). Third: increasingly intricate admixtures of technological and biological components that compute are forcing a redefinition of life itself (Sect 1.3).

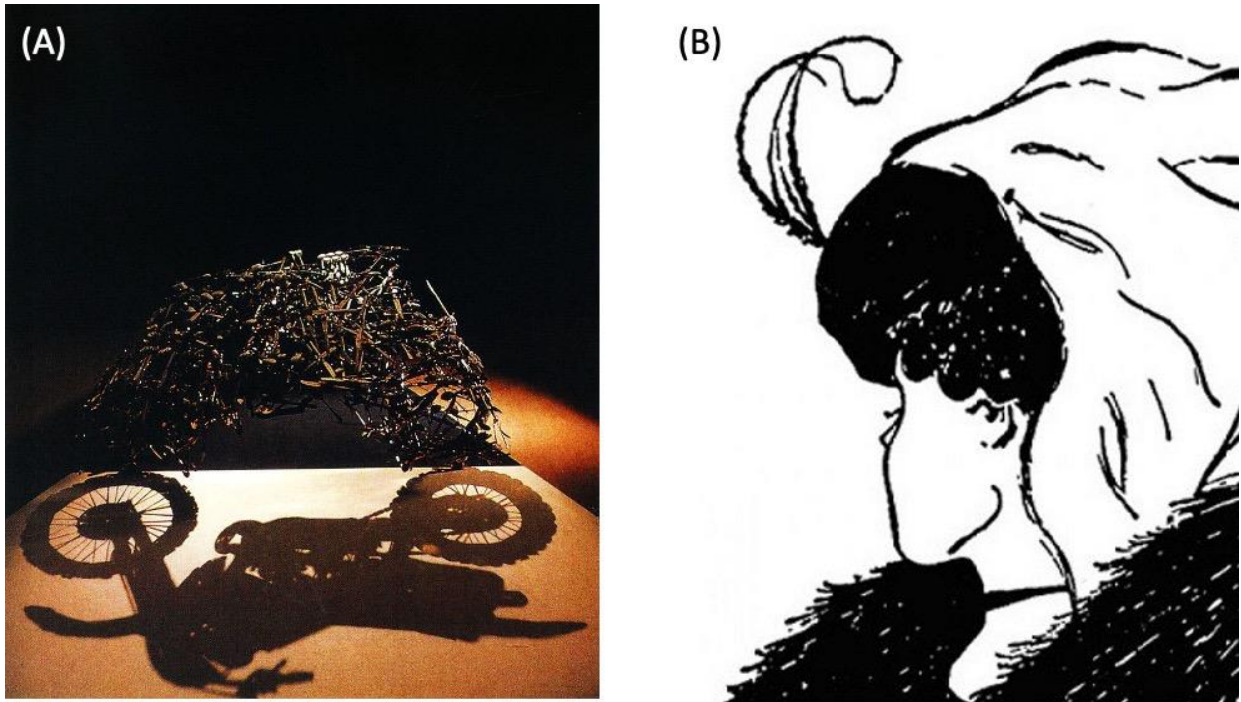

Figure 1: Polycomputing concepts in art.
(A) Sculpture by Shigeo Fukuda, "Lunch with a helmet on", 1987 – appears as a random pile of knives and forks but when observed in just the right way, light moving through the sculpture reveals another pattern (a motorcycle) present at the same time in the same structure.
(B) A well-known bistable (ambiguous) image, "My Wife and my Mother-in-Law" by British cartoonist William Ely Hill in 1915, reveals how our nervous system is not suited to taking in multiple meanings – it prefers to squash down to a single interpretation, even if it then has to vacillate back and forth.

*1.1 What constitutes a computer?* The notion of "computer" needs to be expanded: it no longer only refers to the sequential, deterministic, silicon-embodied, human-programmed, Von Neumann/Turing architectures with which biologists are familiar. Those are indeed not similar to living systems. There is now a widening array of computational substrates and robots that are often massively parallel (such as GPUs and computational metamaterials [8]), stochastic (hard to predict) [9], exploit non-obvious (and potentially not-yet-understood) properties of the exotic substrates they are built from [10], emergent, produced by evolutionary techniques [11], and built by other machines [12] or programmed by other algorithms [13-15]. The benefit of considering biological systems as members of this broader class is that it avails powerful conceptual frameworks from computer science to be deployed in biology in a deep way – to understand life – far beyond the current limited use in computational biology. Moreover, exploring this powerful invariant between natural and synthetic systems can enrich intervention techniques in biology and improve the capabilities of engineered devices, and reveal gaps in our understanding and capabilities in both computer science and biology. Polycomputing is a powerful but as yet under-appreciated example of the many ways in which the wider class of computer devices can help revolutionize the life sciences. In the same way that organic and inorganic materials acting as computers increasingly challenges the claim that living materials are not computers, we have argued elsewhere [13] that the widening array of materials that can now be viewed, or engineered with, as machines is corroding the classic claim that living systems are not machines, and forcing an improved definition of "machine" that escapes the narrow definitions of past decades which are no longer appropriate [14-16].

*1.2 Observer dependency*. In the statement "living things are (or are not) computers", "are" implies the existence of an objective, privileged view of both computers and biology that allows an unambiguous, universal decision as to whether they are related. This binary view is untenable and gives rise to numerous pseudoproblems. We argue instead for an observer-dependent, gradualist view in which computational formalisms are just metaphors, but indeed, *all* scientific concepts are just metaphors, with varying degrees of utility (not binary). Once we come to grips with the fact that "all models are wrong but some are useful" [16], it is possible to adopt a pragmatic approach [17] in which anything is a computer, in a given context, to the degree to which it enables some observer to predict and control that thing better than competing metaphors allow us to do. On this view, whether something is computing is not a philosophical question, but one to be settled experimentally by specifying a computational framework and showing empirically what new levels of capability, and what new experiments and research, are enabled by adopting that framework. Of course, it is expected that future progress will uncover even better ones, so the answer is never final but always provisional and relative to a specific perspective. This view is akin both to the intentional stance in philosophy of mind [18], driving the development of frameworks and tools from cognitive science deployed broadly across biology and the biomedical sciences [2,19,20].

1.3 *What things are alive?* Finally, the question of what constitutes a "living thing" is itself undergoing a renaissance due to the new chimeric, synthetic, and bioengineering techniques being developed [21]. Active matter, synthetic biology, and biohybrids [22-28] are blurring the line between evolved and designed systems, and dissolving distinctions between "life" and "machine" [29-31], which were easy to maintain when our capabilities did not permit the construction and analysis of the full option space of agents [32,33]. At this point, the life sciences have expanded well beyond the N=1 example of phylogenetic history here on Earth, to a clear mandate to understand life-as-it-can be via synthetic and exobiological exploration [34-40].

1.4. *From a philosophy to a science of how life (poly)computes*. We propose that the way to side-step philosophical debates about whether biological systems "are" computers is to adopt an observer-centered, gradualist view of computational formalisms in biology. Polycomputing is an ideal example of a linking concept that will enrich both fields, and which enables a number of fascinating questions with many fundamental and practical implications. What are key functional and control properties of polycomputing systems? How does evolution create systems where multiple functions reside in the same hardware, and what does this design principle mean for evolvability? How can we derive intervention policies to make rational changes in existing polycomputing systems, and what are efficient paths to the design of novel polycomputing materials, control algorithms, and device architectures?

      Regardless of whether or not a living system is distally observed, it still polycomputes because life itself adopts the same operator-dependent approach. In other words, a biological

mechanism is polycomputing because its functionality and signaling are interpreted in different ways by other components of that same living system. Each level and component of a living system is simultaneously an observer and hacker, interpreting and taking advantage of different aspects of mechanisms in its microenvironment, in parallel. Life polycomputes because it is a set of overlapping, competing, cooperating nested dolls each of which is doing the best it can to predict and exploit its microenvironment [41-47].

*1.5. Why "life as computation" matters.* Transfer of knowledge between the disciplines of biology and computation forms a positive feedback loop for increasing insight in both. Biological examples help widen the range of implementations for computing devices and provide novel ideas for architectures [48-52]; unconventional computing platforms include fungal networks, ant colonies, and DNA. In complement, computer science and its ideas of functionalist substrate independence (multiple realizability) help biologists focus on essential, not contingent, design principles – expanding biology beyond zoology and botany. This has been most prevalent in neuroscience [53-55], but more recently has been extended far beyond, in recognition of the fact that neural dynamics are only an extension of far older biological problem-solving architectures [20,56-58].

A key conceptual insight from computer science that informs biology concerns the nature of computation. For example, the field of physical reservoir computing [59], in which a neural network is trained to map dynamics occurring within an inorganic, biological or technological system (the "reservoir") into some output desired by a human observer, helps us see the observer-dependent aspect of biology. This offers ways to think about biology as nested societies of elements which are exploiting the information-processing capacities of their living environment. Cells, parasites, conspecifics, commensal organisms, and evolution itself, are all hackers in the sense of using their parts and their neighbors as affordances in whatever way they can, rather than in some single, unique, privileged objective way that reflects "true" functionality.

The concepts of superposition in quantum mechanics and primacy of observer frames in relativity transformed understanding of phenomena at, respectively, very small and very large scales. Polycomputing challenges us to apply the same concepts to computation and life at meso-scales. Here, we overview the concepts of superposition and observer frame as they apply to meso-scales and argue that the polycomputing lens, like the agential matter lens [60,61], helps us understand, predict, and control new classes of evolved and designed materials, with numerous applications ranging from regenerative medicine to engineering.

## 2. Current Debates: dissolving dichotomous thinking

Whenever technological progress in a particular domain begins to slow, researchers often look to nature for fresh inspiration. Examples include photosynthesis for new energy capture devices [62] and flapping wings for new drone designs [63]. Following in this tradition, the increasing difficulty of packing more compute into microchips [64] challenges us to seek new paths forward by considering how computation is embedded in living systems. Comparing how organisms and machines compute requires one to view an organism as a kind of machine; otherwise, no comparison is possible. The debate about how or whether organisms are machines has a long history, and has become more intense in recent years [29-31,56,65-67] as various disciplines not only compare life to machines, but attempt to merge the two (reviewed in [32]).

Our usage of the term "machine" in what follows will denote that subset of machines capable of computation. Such machines include robots and physical computers but exclude simple mechanical devices such as combustion engines and flywheels, for which no way to stimulate or exploit them to produce computation has yet been invented. (If such interventions are discovered, these technologies can then be considered as belonging more to the class of computational machines.) In the spirit of our thesis, we acknowledge that there is no clear dividing line between these two "types" of machines, as circuitry-free machines like physical reservoir computers [59] and computational metamaterials [91] can still compute to some degree. As always, there is a continuum: in this case, across machines capable of more or less computation. The usage of the term "machine" rather than "computer" in what follows is meant to remind the reader that we are considering organisms vis-a-vis human-made things that compute, rather than just comparing them to traditional computers.

*2.1. Structure function mapping and polycomputing.* An obvious starting point for comparing organisms and computers, or organisms and machines, is to assume a 1-to-1 mapping between structure and function. A comparison can then be attempted between the organism's and machine's structures, and then between their functions. Finally, one can compare the structure-to-function mappings of organisms and machines. However, teasing apart structure and function for such comparisons is difficult. Genetics [68] and neuroscience [69] both provide historical examples of how 1-to-1 structure/function mappings were rapidly replaced by models with increasingly dense and non-intuitive interactions between structural and functional units. Even harder than making predictions based on this nontrivial structure-to-function mapping is the inferring of interventions to make rational changes at the system level, as is needed for example for regenerative medicine – replacing complex organs such as hands and eyes [19,20]. Advances in other areas where biology and computer science meet are similarly demolishing long held dichotomies (Table 1).

Indeed, an understanding of the wide range of implementations (materials, including organic components) and origin stories (e.g., evolutionary design techniques [70]) for machines makes it clear that in many cases, a modern machine lens on life facilitates progress. The machine metaphor is a functional approach that seeks to develop possible efficient ways to predict, control, communicate with, and relate to a system and its reliable behavior modes. However, one aspect has lagged, in both engineering and biology. It is relatively easy to see that technological or living components can support different functions at the same time but at different spatial scales: myosin, for example, supports muscle fiber contraction and legged locomotion simultaneously. It is also easy to see how components can support different functions at the same spatial scale but at different times: myosin can support legged locomotion and then tree climbing. But it can be difficult to see how a component can provide multiple uses to multiple beneficiaries (or compute different functions from the vantage point of different observers) at the same spatial scale and at the same time. Investigating this last phenomenon---polycomputing---enables not only a new set of questions for biology, but also a quest for engineers to understand how to pack more functionality into the same machine.

| Assumed distinction | Counterexamples |
|---|---|
| Software/Hardware | Physical materials that compute [59] and learn [71]. |
| Tape/Machine | Tape-less von Neumann self replicators [72]. |
| Digital/analog | Evolved digital circuits can exploit electromagnetic properties of the circuit's substrate [11]. |
| Machine/Life form | AI-designed organisms [72,73]. |
| Automaton/Free agent | The intentional stance [18]. |
| Brain/Body | Computational metamaterials [8]. |
| Body/Environment | Other cells are the environment for a cell in a multicellular body. |
| Intelligent/Faking it | AI technologies that seem to pass verbal [74], visual [75], or physical [76] Turing tests. |
| Created/Evolved | Artefacts designed by human-created evolutionary algorithms. |

**Table 1**: Some common assumed distinctions in biology and technology, and recent advances that serve as counterexamples, suggesting a spectrum of complementarity.

2.2. *Dichotomous thinking in the life sciences*.

Biology does not really support dichotomous categories. While it is sometimes convenient for biologists to adopt discrete criteria for specific characters, evolution and developmental biology both exhibit remarkable gradualism. Neither process supports any kind of clean bright line that separates the cognitive human being from the "just physics" of a quiescent oocyte or "true grounded knowledge" from the statistically-driven speech behavior of babies and some AIs, etc. (Table 1). All of these, like the process of slowly changing a being from a caterpillar to a butterfly [41], show that familiar categories in fact represent poles of a spectrum of highly diverse mixed properties. The interoperability of life [41,77-79] enables chimeras at all levels of organization, which provide a continuum of every possible combination of features from supposedly distinct categories (Table 1), making it impossible to objectively classify either natural and artificial chimeras [32,80]. It is becoming increasingly apparent that the departmental, funding, and publication distinctions between disciplines (e.g., neuroscience and cell biology), are much more a practical consequence of our cognitive and logistical limitations than the reflection of a deep underlying distinction. In fact, these divisions obscure important invariants – symmetries across categories that enable unifications such as the use of cognitive neuroscience techniques to understand the collective intelligence of cells during morphogenesis [19,20,81,82] or indeed, of physics itself [83,84].

2.3. *Dichotomous thinking in computer science*.

Advances in the computational sciences also increasingly disrespect human-devised categorical boundaries. One such boundary under attack is that between body and brain. One set of technologies eating away at this distinction is physical computing; a conceptual advance doing similarly caustic work is that of morphological computation. In mechanical computing, computation is performed without recourse to electronics and instead relies on optical [85], mechanical [86], or quantum [87] phenomena. Recent advances in mechanical computing show how inert bulk materials can be induced to perform non-trivial calculations including error backpropagation, the algorithmic cornerstone of modern AI [71]. The recent demonstration by one of the authors (Bongard) that propagation of acoustic waves through granular metamaterials can be exploited to perform multiple Boolean operations in the same place at the same time [8], can be considered the first example of mechanical polycomputing. Mechanical computing thus challenges the assumption that the brain (or the circuitry) is the thing that computes (or polycomputes); while the body (or the robot body) is a separate thing that does not, or cannot, compute.

Morphological computation, a concept originating in the robotics literature, upholds that the body of an animal or robot can indeed compute and, moreover, it can "take over" some of the computation performed by a nervous system or robot control policy [88,89]. Although mechanical computing and morphological computing are similar in spirit, in mechanical computing, the bulk material passively accepts whatever computation is forced upon it. In contrast, in morphological computation, the animal or robot body is active: it is capable of taking on computational responsibilities via evolution or learning. This flow of computation back and forth between body and brain (or between digital circuitry and bulk materials) suggests that the two human-devised categories of "body" and "brain" should not be as distinct as once thought.

2.4. *Polycomputing in bodies and brains*.

If polycomputing is to be considered a kind of computation, one can then ask whether polycomputation can be shuttled back and forth between biological bodies and brains, or can be made to do so between machine bodies and brains. For this to work, polycomputation must be implementable in different kinds of substrates. Traditional computation is assumed to be substrate agnostic: if configured appropriately, any physical material can compute. In contrast, only vibrational materials currently seem capable of polycomputing, as polycomputation requires storage of the results of multiple computations at the same place and at the same time, but at different peaks in the frequency spectrum. This would seem to preclude some materials, like digital circuitry and biological nervous systems, from performing polycomputation, since digital circuitry traffics in electrons, and nervous systems traffic in chemicals and ions; neither seems to traffic in vibration. At first glance, this seems poised to rescue the brain/body distinction via the surprising route of suggesting that bodies and brains are different things because bodies polycompute but brains do not.

However, this odd-seeming distinction may be short-lived. It has been shown that neurons may communicate mechanically [90] in addition to electrically and chemically. If so, such mechanical neural communication may contain vibrational components, suggesting nervous systems may be polycomputing as well. If this turns out to be the case, it in turn opens the possibility that nervous tissues may have evolved incremental enrichments of non-neural cells' already proven ability to polycompute. This would once again frustrate our attempts to cleave body from brain, in this case by the claim that one polycomputes, while the other does not.

Mechanical computing and morphological computation are closely related to another way in which computer science provides useful viewpoints for biology. In CS, the view that an algorithm *drives* (functionally determines) outcomes, even though implemented by the microphysics of electron flows through a CPU, is accepted and indeed essential to perform the useful and powerful activity of programming. This is in stark contrast to debates in biology and neuroscience about whether higher levels of description are merely epiphenomenal [91-95], supervening on biochemical microstates (reductionism). Computer science clearly shows how taking a causal stance at higher levels enables progress. Indeed, the recent advances in information theory around quantifying causal emergence [95,96] show how the same Boolean network can be computing different functions simultaneously (Figure 2), depending on the level of analysis chosen by an observer [95]. This has interesting biological implications, since such networks are a popular model for understanding functional relationships between genes [97-99].

Biological nervous systems - the human brain in particular - have attracted increasingly computational metaphors throughout the industrial revolution and information age. The application of computational concepts to brains has had unintended consequences, most of all the implicit assumption that tissues, cells, and other biological systems that are not brains do not compute. However, the brain:body dichotomy is increasingly being dismantled by studies of basal cognition (i.e., intelligence in unfamiliar embodiments) in plants [100,101], single somatic cells [102-104], microbes [105-108], and at the tissue level in organisms [109-112]. Indeed, the bioelectric and neurotransmitter dynamics that implement predictive processing and other computations in brains are speed-optimized versions of extremely ancient bioelectrical computations that navigated spaces (such as anatomical morphospace, physiological space, etc.) long before brains and muscles appeared [58,113,114]. Indeed, the tools of neuroscience – from conceptual approaches such as active inference [20,82] to the molecular tools like optogenetics [115-118] – do not distinguish between neurons and non-neural contexts, being applicable broadly across biology.

The benefit of dissolving these arbitrary distinctions is that commonalities and fundamental design principles across substrates are beginning to emerge across evolved and designed constructs at all scales [32,103,119]. Frameworks that are to survive the next decades, in which technological advancement will further enmesh biology and technology, must facilitate experimental progress at the expense of philosophical preconceptions. More than that, they must provide unifying insight by identifying symmetry and deep order across fields, to combat the ever-growing problems of big data and the interpretability crisis [120,121]. Here, we delve into one emerging principle: polycomputing, which places front and center the fascinating issues of form, function, control, interpretation, and the role of the observer.

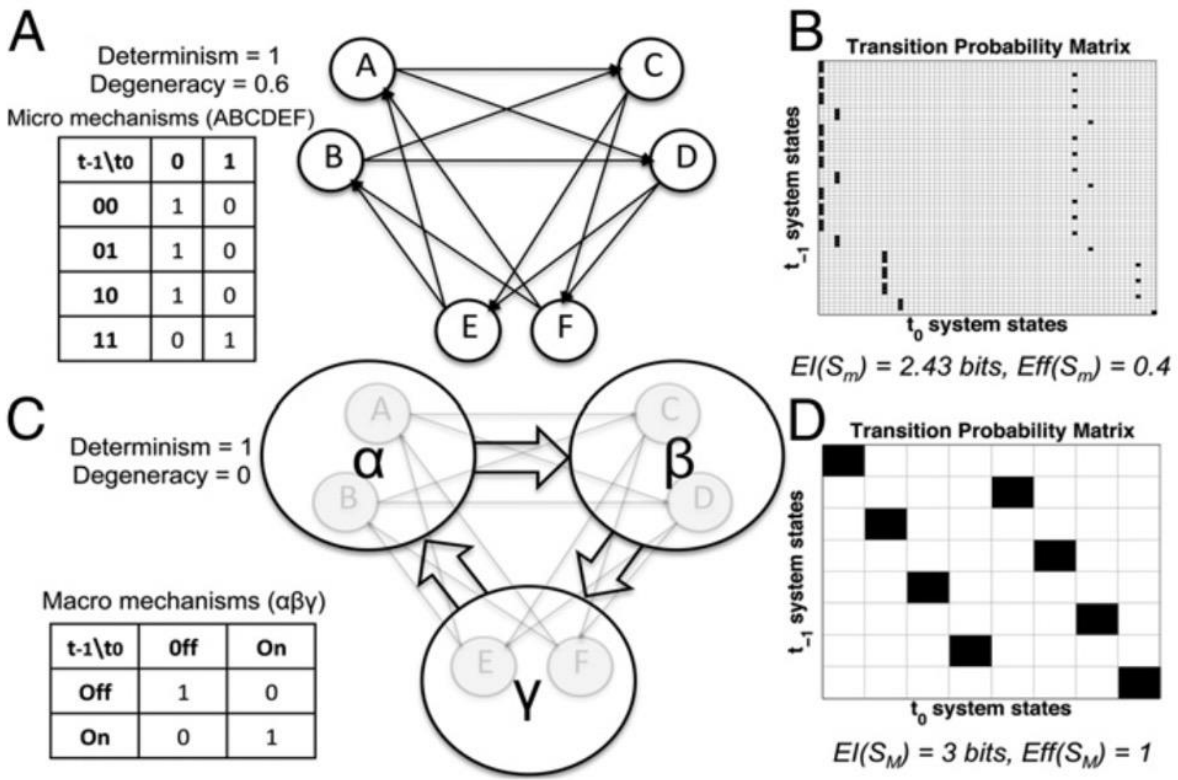

Figure 2: Spatial causal emergence (counteracting degeneracy).
(A) A degenerate network with deterministic AND gates.
(B) The cycle of AND gates is mapped onto a cycle of COPY gates at the macro level.
(C) The deterministic but degenerate micro TPM.
(D) The deterministic macro TPM with zero degeneracy.
By eliminating degeneracy and achieving perfect effectiveness, the macro scale beats the micro scale in terms of actionable information content (causal emergence = 0.57 bits). This system is simultaneously computing a set of ANDs and, at a macroscale, a set of COPY gates as well, depending on the level of observation. Taken with permission from [95].

## 3. Learning from superposed systems in engineering

The ability to compute and store multiple results in the same place at the same time is of obvious engineering interest as it could greatly increase computational density. Various technologies are now being built that roll back the assumption that such superposition is impossible. These technologies, reviewed below, suggest ways to look for similar phenomena in natural systems.

Quantum computing has made clear that multiple computations can be performed simultaneously. But practical and general-purpose quantum computing remains a distant prospect. Recently, one of the authors (Bongard) has shown that quantum effects are not necessary for polycomputing [8]: Even relatively simple materials composed of only 30 parts are capable of computing two logical functions in the same place at the same time. This non-quantum form of computational superposition suggests not only that more computation may be packed into smaller spaces, but also that the fundamental way in which computation arises in technological and biological materials may need to be rethought.

Holographic data storage (HDS; [122]) is another set of related technologies that do not assume only one datum or computational result is stored locally. HDS stores and reads data that has been dispersed across the storage medium. It does so by etching a distributed representation of a datum across the storage medium, for example with laser light, from a particular direction. That datum can then be retrieved by capturing the reflection of light cast from the same direction. By storing data in this way from multiple directions, parts of multiple pieces of data are stored in the same place, but accessed at different times. Exactly how this can be achieved in hardware such that it affords appreciable increases in storage density over current traditional approaches has yet to be resolved.

A third technology relaxing the assumption of data/compute locality is physical reservoir computing (PRC). PRC, inspired by HDS, attempts to retrieve the results of desired computations by exciting inert bulk materials, such as metal plates or photonic crystals, and capturing the resulting vibrations or refracted light, respectively. Different computations can be extracted from the same material by exciting it in different ways. An attempt to "program" PRCs, thus easing the ability to extract desired computation from them, has also been reported [123]. Notably, this method has been used to create "deep physical neural networks" [71]: the input, and parameters describing an artificial neural network, are combined into forces supplied to the material. Forces captured back from the material are interpreted as if the input had been passed through a neural network with those parameters. Errors in the output can then be used to modulate the input, and the process repeats until a set of input forces has been found that produces the desired output. Importantly, the internal structure of the bulk material is not changed during this training process. This means that the same material can embody different computations. Just how distributed or localized these computations are within these materials remains to be seen.

Other materials are not only changed by forces acting on them, but retain an imprint of those forces even after they cease; they are capable of memory. Efforts are also underway to design such materials to maximize the number of overlapping memories that they can store [124,125]. The ability of designed materials to absorb forces, compute with them, and output transformed forces encoding the results of those computations holds great promise for robotics. If future robots can be built from such materials, force propagation within them could simultaneously produce external behavior and internal cogitation, without requiring distinct behavior-generating components (the body) and computation-generating components (the brain). Indeed, soft robots are already demonstrating how exotic materials enable the traditionally distinct functions of sensation, actuation, computation, power storage, and power generation to be performed simultaneously by the same parts of the robot's body [126].

**4. Biology is massively overloaded: polycomputing**

Analyzing natural systems to determine whether or how they perform polycomputation is particularly challenging, as most analytic approaches are reductionist: they "reduce" to characterizing one phenomenon that arises in one place, at one time, under one set of circumstances. Synthesis is also difficult: polycomputable technologies seem, to date, resistant to traditional engineering design principles such as hierarchy and modularity. A fundamental problem is that typical human designs are highly constrained, such that changes made to optimize one function often interfere with another. Although humans struggle to manually design polycomputing technologies, it turns out that AI methods can do so, at least in one domain. We have recently applied an evolutionary algorithm—a type of AI search method—to automatically design a granular metamaterial that polycomputes. It does so by combining vibrations at different frequencies at its inputs, and providing different computations in the same place, at the same time, at different frequencies. Figure 3 illustrates this process.

Many biological functions have been usefully analyzed as computations [56] (Table 2). These include molecular pathways [124,127], individual protein molecules [57], cytoskeletal elements [125,128,129], calcium signaling [130], and many others. Although differing from the most familiar, traditional algorithms, the massively parallel, stochastic (indeterministic), evolutionarily-shaped information processing of life is well within the broad umbrella of computations familiar to workers in the information sciences. Indeed, information science tools have been used to understand cell- and tissue-level decision-making, including estimations of uncertainty [104,131-141], analog/digital dynamics [142], and distributed computation [100]. Bioelectric networks within non-neural tissues, just like their neural counterparts, have shown properties very amenable to polycomputation, including ability to store diverse pattern memories that help execute morphogenesis on multiple scales simultaneously [20,22,112,143-149], and enables the same genome to produce multiple diverse outcomes [150].

A key aspect to recognizing unconventional computing in biology is that the notion of "what is this system *really* computing" has to be dropped (because of the multiple observers issue described above). Once we do this, biology is rife with polycomputing at all scales. Examples include the storage of a (possibly uncountable) number of memories in the same neuronal real-estate of the brain [151,152], and many others summarized in Table 2. We do not yet know whether the prevalence of polycomputing is because of efficiency, robustness, or other gains that override the evolutionary difficulty of finding such solutions. Or, perhaps we overestimate the difficulty, and evolution has no problem identifying such solutions – they may indeed be the default. If so, it may be because of the generative, problem-solving nature of developmental physiology as the layer between the genotype and the phenotype [32,153]. For these reasons, polycomputing may join degeneracy and redundancy [154] as one of the organizing principles that underlies the open-ended, robust nature of living systems.

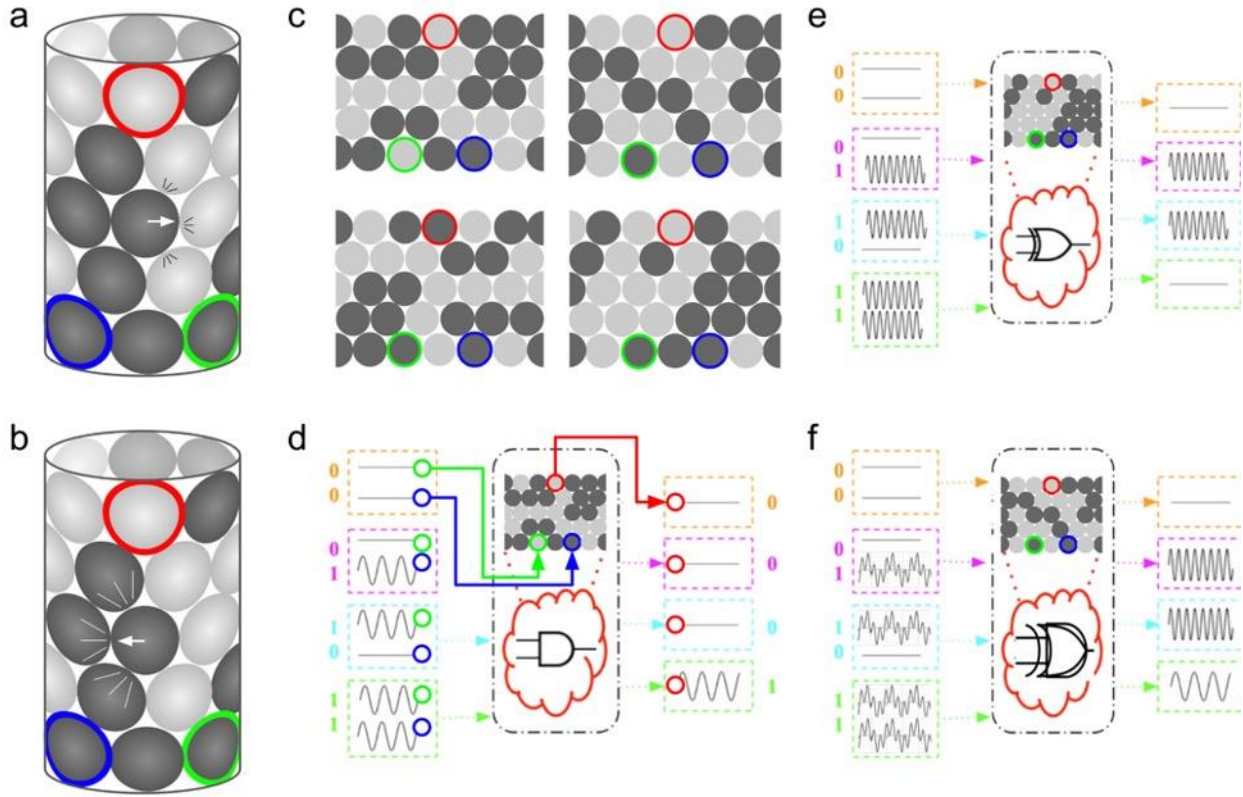

Figure 3: Engineering polycomputing materials.
(A) A granular metamaterial can be assembled by embedding stiff (dark gray) and soft (light gray) particles into a sheet and wrapping it into a tub. If a particle collides with soft particles, it only affects their motion a bit.
(B) if it hits rigid particles, their motion is affected more.
(D) An evolutionary algorithm (EA) can be created that evolves populations of metamaterials, where each one has a unique combination of stiff and soft particles. The EA can then delete those metamaterials that perform poorly at some desired task, such as performing a computation, and make randomly modified copies of those that do a better job.
(D) This can result in the evolution of a material that acts as an AND gates, a building block of computers: some designated 'output' particle (red) should only vibrate if two other 'input' particles are vibrated from outside the system (green and blue).
(E) An evolutionary algorithm can be asked to evolve a metamaterial that acts as an AND at one frequency, but to act as another computational building block, an XOR gate, at a higher frequency: the output particle should only vibrate if one of the input particles is vibrated.
(F) This process results in the evolution of a polycomputing material: if inputs are supplied at two different frequencies, the evolved material acts as an AND and XOR gate simultaneously: it provides the results of these two computations at the same place at the same time (the output particle), but at different frequencies. Details can be found in [8].
14

Table 2:  Examples of biological polycomputing at diverse scales

| Multiple Computations in the Same Biological Hardware | Reference |
|---|---|
| Mitochondria also act as micro-lenses in photoreceptors | [155] |
| Proteins acting in multiple (fluctuating) conformations | [156] |
| Gene regulatory networks with multiple memories/behaviors | [157-160] |
| Chemical networks performing neural network tasks | [161,162] |
| RNA encoding enzyme and protein functions | [163-166] |
| DNA with more than one active reading frame (overlapping genes) | [167,168] |
| Ion channels that are also transcription factors | [169] |
| DNA transcription factors working in DNA replication machinery | [170] |
| Polysemanticity and superposition of meaning in neural networks and language understanding | [171-173] |
| Cytoskeleton performing computations via simultaneous biomechanical, bioelectrical, and quantum-mechanical dynamics | [174-183] |
| Electrophysiological networks performing memory functions while regulating heartbeat | [184] [185,186] |
| Bioelectric networks performing physiological functions while also regulating morphogenesis | [113] |
| Spider webs as auditory sensors and structural elements | [3] |
| Pleiotropy – most genes have multiple functions | [68] |
| Holographic memory in the brain | [187] |
| Multiple behaviors in the same neuronal circuit | [188] |
| Multiple personalities in the same brain (dissociative identity disorder and split brain studies) | [189,190] |
| Calcium dynamics performing as a hub in a huge bowtie network of diverse simultaneous processes | [191,192] |

4.1. *Evolutionary pivots: origins of polycomputing?* Evolution is remarkably good at finding new uses for existing hardware because of its fundamental ability to generate novelty to exploit new niches while being conservative in terms of building upon what already exists. This ability to simultaneously innovate and conserve plays out across structural, regulatory, and computational domains. Moreover, re-use of the same conserved mechanisms *in situ* enabled evolution to pivot successful algorithms (policies) from solving problems in metabolic space to solving them in physiological, transcriptional, and anatomical (morphospace) spaces, and finally, once muscles and nerves came on the scene, to 3D behavioral spaces [193]). For example [58], the same ion channels that are used for physiological control of cell homeostasis and metabolism are used *simultaneously* in large-scale bioelectric circuits that compute the direction of adaptive changes in growth and form in embryogenesis, metamorphosis, regeneration, and cancer suppression [58,108,113,194-198]. Indeed, in some animals such as planaria and axolotl, this is all happening at the same time as these exact same mechanisms in neural cells are guiding behavior [199]. Biology is rife with polycomputing because it uses a multi-scale competency architecture – every level of organization is competent to solve certain problems in its own space and is doing so at the same time via the same physical medium [2,200] (Figure 4).

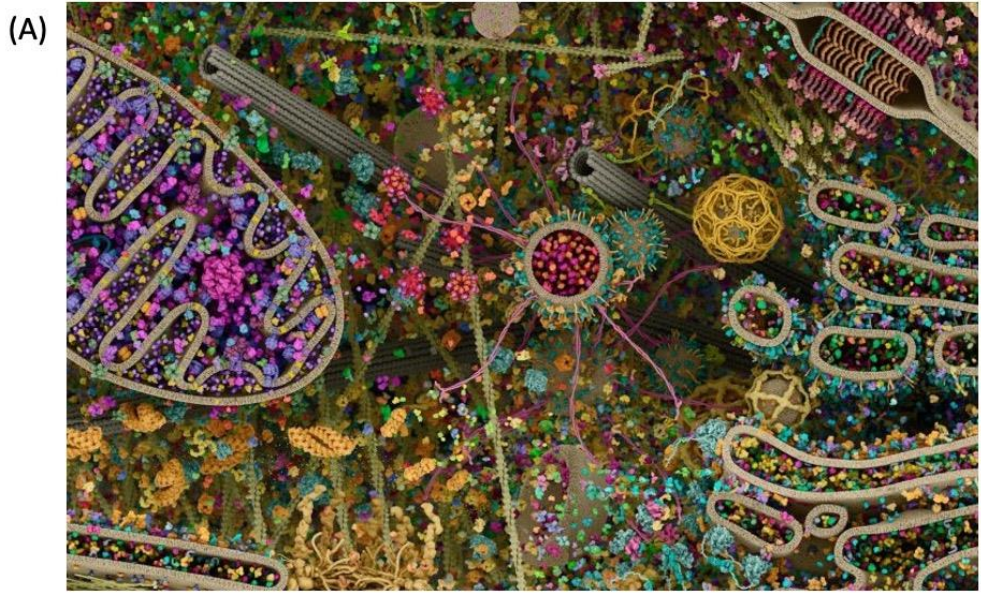

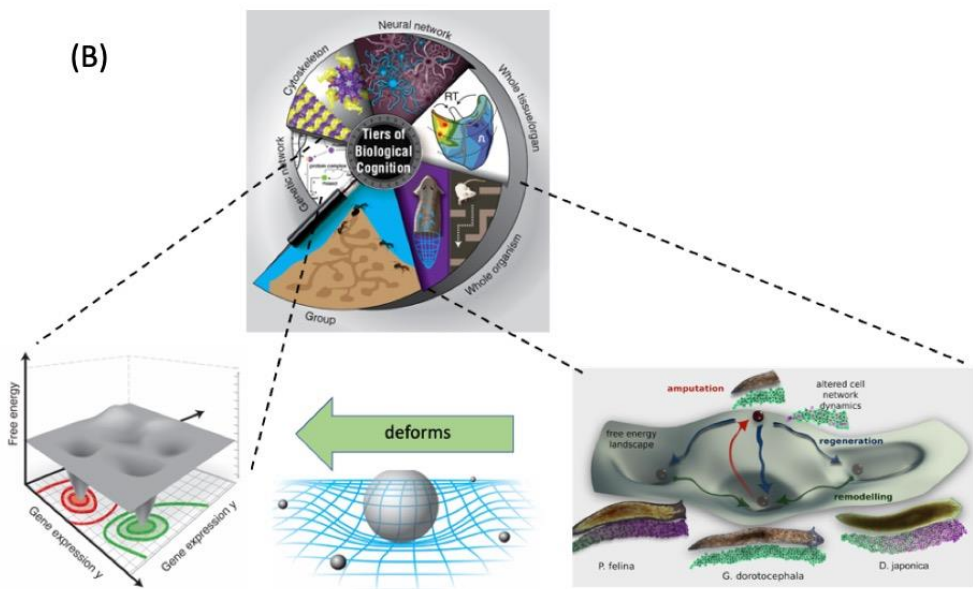

Figure 4: Polycomputing architectures in biology
(A) Gael McGill's computer model of the inside of a cell, combining all of the known components that occupy the same space. This illustrates the pressure on biology to have each component do multiple duty (there is not much room to add additional components). Used with permission.
(B) Multiscale competency architecture of life consists of molecular networks which make up cells which make up tissues which make up organs which make up organisms within swarms. Each layer is performing specific functions simultaneously; for example, the tissue layer is attempting to compute the correct attractor for the collective morphogenetic behavior of planarian fragment cells, which can build one of several head shapes). Each layer deforms the action landscape for the layer below it, providing incentives and shaping geodesics that force the lower-level components to use their behaviors in service of the higher level's goals. Taken with permission from [200]. Images by Jeremy Guay of Peregrine Creative Inc. and Alexis Pietak.

Polycomputing is seen even at the lowest scale of molecular biological information. It has long been known that genomes are massively over-loaded, providing polycomputing not only because of multiple reading frames (overlapping genes) for some loci [167,168], but also because the question of "what is this gene for?" may have a clear answer at the molecular scale of a protein, but often has no unique answer at the phenotypic scale because complex traits are implemented by many genes, and many (or most? [68]) genes contribute to multiple mesoscale capabilities. Moreover, epigenetics enables the same genomic information to facilitate the embodied computation that results in multiple different anatomical, physiological, and behavioral forms [201,202].

*4.2. Polycomputing and the range of phenotypic possibility.* How much actionable information on how to assemble an adaptive, behaving organism can be packed into the same genomic and physiological information medium? A recent example of massive phenotypic plasticity are the Xenobots – proto-organisms that result from a rebooting of multicellularity by frog skin cells [73,203]. The Xenobots self-assemble as spheroids that are self-motile, exhibiting a range of autonomous behaviors, including highly novel ones such as kinematic self-replication: the ability, of which Von Neumann famously dreamed, to assemble copies of themselves from material found in their environment [72]; in the case of Xenobots, dissociated cells introduced into their surroundings. A key point is that Xenobots are not genetically modified – their novel functionality is implemented by perfectly standard frog cells. So, what did evolution learn [204-208] in crafting the *Xenopus* laevis genome and the frog egg's cytoplasmic complement? It was not just how to make a frog; it is how to make a system in which cells allow themselves to be coerced (by other cells) into a boring, 2-dimensional life as the animal's outer skin surface, or, when on their own, to assemble into a basal 3-dmensional creature that autonomously explores its environment and has many other capabilities (Figure 5).

This capacity to do things that were not specifically selected for [209] and do not exist elsewhere in their (or others') phylogenetic history reveals that evolution can not only create seeds for machines that do multiple things, but ones that can do *novel* things. This is because genomic information is overloaded by physiological interpretation machinery (internal observer modules): the exact same DNA sequence can be used to build a tadpole or a Xenobot (and the same planarian genome can build the heads of several different species [210,211]).  Thus, evolution teaches us about powerful polycomputing strategies because it does not make solutions for specific problems – it creates generic problem-solving machines, in which the competition and cooperation of overlapping, nested computational agents at all levels exploit the ability of existing hardware to carry out numerous functions simultaneously. This is closely tied to recent advances at the interface of the fields of developmental biology and primitive cognition, with the generation of models in which larger-scale Selves (in psychological, anatomical, etc. spaces) arise as composite systems made of smaller Selves, all of which are pursuing diverse agendas [199,212,213].

As surprising as these examples are, we should have already seen this coming – biology *had to* work like that, and it could not work otherwise. First, the "sim-to-real gap" (the difference an agent experiences when trained in virtual environments, built as a robot, and deployed into a real environment [214,215]) is as real for biology as it is for robotics: prior evolutionary experience in a past environment is not a reliable guide to the novel challenges each generation experiences in new environments. Thus evolution does not overtrain on prior examples, but generalizes, producing substrates that can compute different functions for different needs (Figure 6). Second, the evolutionary process is not working with a blank slate – the canvas of embryogenesis is made of cells, which used to be independent organisms. Evolution exploits the competencies of cells in numerous problem spaces as a toolkit of affordances to be exploited. Development is akin to behavior-shaping, where evolution finds signals that cells can send to other cells to push them into specific actions. This is a strong start for polycomputing as a design principle – working with an agential material [60] requires strategies that don't establish a single, privileged new way of doing things but instead drive adaptive outcomes by managing the many things that the material is already doing. Third, evolution simply wouldn't work well with an architecture that didn't support polycomputing, because each new evolutionary experiment would wreck prior gains, even if it itself was an advance.

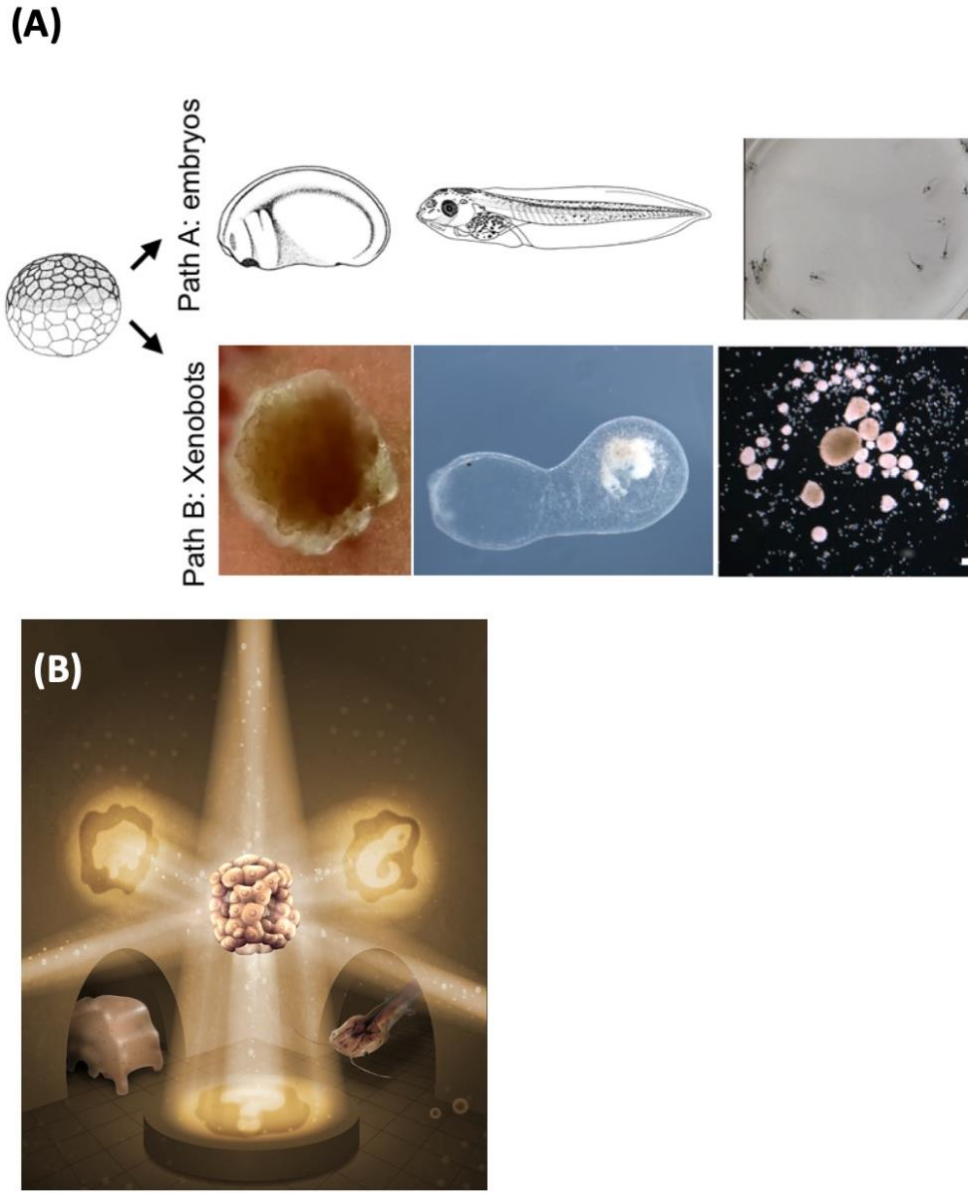

Figure 5: What does the Xenopus laevis genome specify?
(A) The standard Xenopus laevis genome (in a frog egg) typically causes the construction of a set of standard embryonic stages (Path A) which results in tadpoles with specific behaviors. However, in a different context, the skin cells can autonomously create a Xenobot (Path B) – a spherical construct with autonomous motion, a different morphogenetic sequence, and behaviors such as kinematic self-replication. The same genomic information contains simultaneously seeds of emergent tadpoles or Xenobots.
(B) Similar to the iconic cover image of the classic book Godel, Escher Bach [245], this image illustrates how the same hardware (the standard frog egg in the middle) can be used to generate diverse forms of living constructs. Different environments, external signals, and physiological events can coax diverse morphogenetic outcomes out of a constant biological information string (DNA). Images in A courtesy of Xenbase and Douglas Blackiston, Levin lab. Image in B by Jeremy Guay of Peregrine Creative.

*4.3. Evolving polycomputing.* The rate of evolution would be much slower without this multiscale competency architecture – the ability of parts to get their job accomplished even if circumstances change [216]. In one remarkable example, the tadpole eye, placed in the wrong position on the head or even on the tail [217] still provides vision, because the eye primordia cells can make an eye in aberrant locations, move it if possible [218] and if not, connect to the spinal cord (rather than directly to the brain), providing visual signals that way [219,220]. This competency of the substrate in regulative development and remodeling [32,200] can neutralize the lethal side effects of many mutations, enabling exploration of other possibly beneficial effects. For example, consider a mutation that causes displacement of the mouth and also another effect, *E*, elsewhere in the body. The potential benefits of *E* might never be explored in a monocomputational developmental architecture because the mouth defect would prevent the animal from eating and drive fitness to 0. Exploration of the effect of *E* would have to wait for another mutation to appear that produces the same effect without untoward side effects elsewhere – a very long wait, and often altogether impossible. In contrast, in a polycomputing architecture, structures solve morphological and physiological problems simultaneously: the mouth will move to the right location on its own [218], in parallel to all of the other developmental events, enabling evolution to explore the consequences of *E*. Thus, the overloaded competencies of cells and tissues allow evolution to simultaneously explore other effects of those mutations on phenotype (pleiotropy).

In this case, these competencies create the "hidden layer" of developmental physiology that sits between genomic inputs and phenotypic outputs and provides problem-solving capacity: getting an adaptive task completed despite changes in the microenvironment or in their own parts [2,200]. This occurs simultaneously at all scales of organization (Figure 4) and thus each level computes specific functions not only in its own problem space, but also participates in the higher level's space (as a component) and has an influence that deforms the action space of its lower levels' components [200]. By behavior-shaping competent subunits as agential materials [221], evolution produces modules that make use of each other's outputs in parallel, virtually guaranteeing that the same processes are exploited as different "functions" by other components of the cell, the body, and the swarm.

The evolutionary pressure to make existing materials do multiple duty is immense. However, much remains to be learned about how such pressure brings about polycomputing: how some materials can be overloaded with new functions without negatively impacting existing

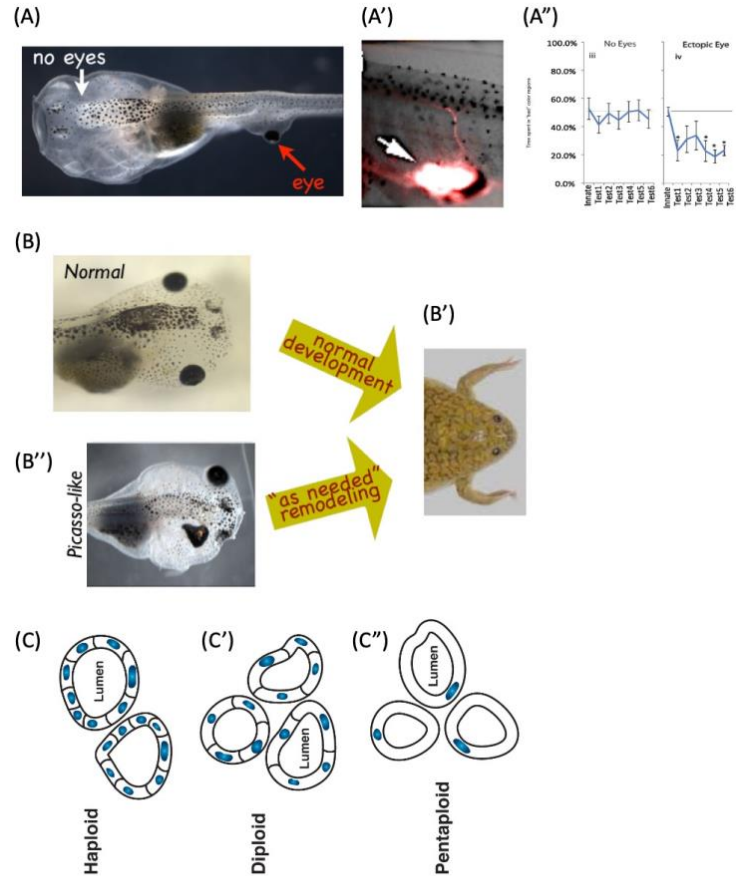

Figure 6: Morphogenesis plays the hand it is dealt

The essence of biological morphogenesis is that it does not assume much about current circumstances and attempts to create a viable organism with whatever is at-hand. Thus, frog embryos in which eye primordia cells are moved from the head to the tail still make good eyes (A), try to connect to the spinal cord (A', red stain), and enable the organism to exhibit behavioral vision (A") despite a completely novel visual system-brain architecture which had no evolutionary prep time to get used to the new arrangement – nothing needed to be changed (at the DNA level) to make this new configuration workable. Similarly, tadpoles (B) which must rearrange their face to turn into a frog (B') can still do so even if everything is artificially placed in a scrambled configuration (B") because each organ is able to move as needed to get its job done (reach a specific region of morphospace). Finally, the cross-level nature of this overloading of basic mechanisms is seen in newt kidney tubules schematized here in cross-section. While they normally consist of 8-10 cells that communicate to make a tubule, the cells can be made experimentally very large – in that case, fewer cells will work together to make the same size tubule (C'). In the case of enormous cells, a completely different mechanism (cytoskeletal bending) will be used by a single cell to create a lumen – showing how the same machine (genome+cell) can enable high-level anatomical goals to trigger diverse low-level molecular mechanisms as needed. Panels A,A',A",B courtesy of Douglas Blackiston, used with permission after [217] and [219]. Panel B' courtesy of Erin Switzer, B" taken with permission from [116]. Panels C,C',C" by Jeremy Guay of Peregrine Creative.

functions. In parallel to such biological investigations, in the computer science domain, much work remains to be done to devise optimization pressures that create polycomputing substrates, and then create new programming strategies suitable for polycomputing. For example, no programming language has yet been devised to truly take advantage of the polycomputational metamaterials described above. Despite our ignorance about how evolutionary or optimization pressures can create polycomputational systems, what is clear is that evolution would not work at all if living things were not machines – predictable, tractable systems. The key aspects of machines are that they harness the laws of physics, computation, etc. in a reliable, rational manner to produce specific, useful outcomes. Evolution exploits the fact that life is a machine by making changes to the material, the control algorithm, and indirectly, to their environment, in a way that gives rise to predictable, adaptive outcomes. Cells could not influence each other during development to reliably achieve the needed target morphologies if they could not be efficiently controlled. Life shows us the true power of the "machine": a powerful multi-scale polycomputing architecture, in which machines control and comprise other machines, all working at the same time in different spaces, produces massive amounts of plasticity, robustness, and novelty.

*4.4 A new approach to identifying and harnessing computational capabilities in vivo and in silico.* One way to exploit this property is to use protocols that examine a particular mechanism for novel things it can do, and for the best way to get it to execute some of its capabilities. At the molecular level, an example is gene regulatory networks (GRNs) – a formalism whereby a set of genes up- and down-regulate each other's function [222,223]. While GRNs and protein pathways are normally studied for ways to explain a particular aspect of biology (e.g., neural crest tissue formation or axial patterning in development [224,225]), we asked whether existing neural network models could have novel computational functions, specifically learning. Our algorithm took biological GRN models and for each one, examined each possible choice of triplets of nodes as candidates for conditioned and unconditioned stimuli and response, as per Pavlovian classical associative learning [226]. We found numerous examples of learning capacity in biological networks and many fewer in control random networks, suggesting that evolution enriches for this property [157]. Most strikingly, the same networks offered *multiple different* types of memory and computations, depending on which of the network's nodes *the observer* took as their control knobs and salient readout in the training paradigm. This approach is an example of searching not for ways to re-wire the causal architecture of the system for a desired function, but searching instead for a functional perspective from which an *un-modified* system already embodies novel functions.

This illustrates an important principle of biological polycomputing: evolution prepared a computational affordance (the GRN) with multiple interfaces (different gene targets) through which engineers, neighboring cells, or parasites can manipulate the system to benefit from its computational capabilities. We suggest that this kind of approach may be an important way to understand biological evolution: as a search for ways in which body components can adaptively exploit other body components as features of their environment – a search for optimal perspectives and ways to use existing interfaces. At the organism level, an excellent example is the brain, in which an immense number of functions are occurring simultaneously. Interestingly, it has been suggested that the ability to store multiple memories in the same neuronal real-estate is implemented by phase [5].

The results of our probing neural networks for novel functions also suggest that alongside tools for predicting ways to rewire living systems [227-229], we should be developing tools to identify optimal perspectives with which to view and exploit existing polycomputing capacities.

**5. Conceptual transitions**

To develop such tools, we will need to overcome human cognitive bias and resist the temptation to cleave phenomena apart in ways that feel comfortable. One approach is to look for particularly non-intuitive phenomena that defy our attempted categories. Better yet is to seek gradients along which we can move from "obvious" approximations of phenomena to increasingly "non-obvious", but more accurate, reflections of reality.

5.1. Directions of conceptual travel.

One such gradient is the one that leads from serial to parallel to superposed processes. The industrial revolution demonstrated the advantage of performing tasks in parallel rather than serially; the computer age similarly demonstrated the power of parallel over serial computation. One reason for these slow transitions may be cognitive limitations: Despite the massive parallelism in the human brain, human thinking seems to proceed mostly, or perhaps completely [230], in a serial fashion. "Traditional" parallelism, as it is usually understood, assumes that multiple processes are coincident in time but not in space. Even more difficult a concept to grasp is that of superposition: the performance of multiple functions in the same place at the same time.

Another conceptual direction that leads from obvious into non-obvious territories is that which leads from modular processes into non-modular ones. The cardinal rule in engineering in general, and software engineering in particular, is modular design. But, this is a concession to human cognitive limits, not necessarily "the best way to do things": many natural phenomena are continua. Taking another step, if we consider biological or technological polycomputing systems, we might ask whether they are modular. But if a system polycomputes, different observers may see different subsets of functions and, some may be more modular than others. In that case, the question of whether a given polycomputing biological system (or bioinspired technology) is more or less modular becomes ill-defined. We argue that to facilitate future research, these classical distinctions must now be abandoned (at least in their original forms).

*5.2. Practical implications for AI/robotics*. Learning how biological systems polycompute, and building that learning into technology, is worth doing for several practical reasons. First, creating more computationally dense AI technologies or robots may enable them to act intelligently and thus do useful, complex work using fewer physical materials and thus creating less waste. Second, technological components that polycompute may be more compatible with naturally polycomputing biological components, facilitating the creation of biohybrids. Third, creating machines that perform multiple computations in the same place at the same time may lead to machines that perform different functions in different domains – sensing, acting, computing, storing energy, and releasing energy – in the same place at the same time, leading to new kinds of robots. Fourth, polycomputing may provide a new solution to catastrophic interference, a ubiquitous problem in AI and robotics in which an agent can only learn something new at the cost of forgetting something it has already learned. A polycomputing agent might learn and store a new behavior at an underutilized place on the frequency spectrum of its metamaterial "brain" better than a polycomputing-incapable agent that must learn and incorporate the same behavior into its already-trained neural network controller. Such ability would be the neural network analogue of cognitive radio technologies, which constantly seek underutilized frequency bands from which to broadcast [231].

**6. Gradual computing in biology: when does the (digital) soul enter the (analog) body?**

The importance of continuous models (and the futility of some binary categories) is readily apparent when tracking the gradual emergence of specific features that we normally identify in their completed state. Examples include pseudoproblems like "when does a human baby become sentient during embryogenesis", "when does a cyborg become a machine vs. organism?", and "when does a machine become a robot?"; all of these force arbitrary lines to be chosen that are not backed up by discrete transitions. Developmental biology and evolution both force us to consider gradual, slow changes as essential to the nature of important aspects of structure and function. This biological gradualism has strong parallels in computer science. An unfertilized human oocyte, mostly amenable to the "chemistry and physics" lens, eventually transforms into a complex being for whom behavioral and cognitive (and psychotherapeutic) lenses are required. What does the boot-up of a biologically-embodied intelligence consist of? What are the first thoughts of a slowly developing nervous system? One key aspect of this transition process is that it involves polycomputing, as structural and physiological functions become progressively harnessed toward new, additional tasks for navigating behavioral spaces in addition to their prior roles in metabolic, physiological, and other spaces [200].

Similarly, one can zoom into the boot-up process when a dynamical system consisting of electrical components becomes a computer. During the first few microseconds when the power is first turned on, the system becomes increasingly more amenable to computational formalisms in addition to the electrodynamics lens. The maturation of the process consists of a dynamical mode which can profitably be modeled as "following an algorithm" (taking instructions off a stack and executing them). Similarly, one could observe externally supplied vibrations spreading through a metamaterial and consider when it makes sense to interpret the material's response as a computation, or the running of an algorithm. In essence, the transition from an analog device to a computer is really just a shift in the relative payoffs for two different formalisms from the perspective of the observer. These are readily missed, and an observer that failed to catch the ripening of the computational lens during this process would be a poor coder indeed, relegated to interacting with the machine via Maxwell's laws guiding electron motion and atomic force microscopy, vs. exploiting the incredibly rich set of higher-level interfaces that computers afford.

6.1. *Agency and persuadability: implication for polycomputing.* One of the most important next steps, beyond recognizing the degree to which certain dynamical systems or physical materials can be profitably seen as computational systems, is to observe and exploit the right degree of agency. Systems vary widely along a spectrum of persuadability [2] – the range of techniques suitable for interacting with them, including physical rewiring, setpoint modification, training, and language-based reasoning. Animals are often good at detecting agency in their environment, and for humans, theory of mind is an essential aspect of individual behavior and social culture. Consistent with the obvious utility of recognizing agency in potential interaction partners, evolution primed our cognitive systems to attribute the intentional stance quite readily [232,233]. Crucially, making mistakes by overestimating agency (anthropomorphizing) is no worse than underestimating agency – both reduce the effectiveness of adaptive interactions with the agent's world.

*6.2. The impact of observer frames.* So far we have considered a single human observer of a biological or technological system, how much agency she detects in the system from her perspective, and how she uses that knowledge to choose how to persuade it to do something. However, a biological system may have many observers (neighboring cells, tissues, conspecifics, parasites) trying to "persuade" it to do different things, at the same time. (Scare quotes here remind us that we must in turn decide to adopt the intentional stance for each of the observers.) A polycomputing system may be capable of acceding to all of these requests simultaneously. As a simple example, an organism may provide a computational result to one observer while also providing waste heat produced by that computation to a cold parasite. Traditional computers are not capable of this, or at least are not designed to do so; but future polycomputational machines might be.

6.3. *Becoming a computer*

For many outside the computational sciences, "computer" denotes the typical physical machines in our everyday lives, such as laptops and smartphones. Turing, however, provided a formal definition for computers that is device-independent: in summary, a system is a computer if it has an internal state, if it can read information from the environment in some way, update its behavior based on what it read and its current state, and (optionally) write information back out into the world. This theoretical construct has become known as a Turing Machine; any physical system that embodies it, including organisms, is formally referred to as a computer. This broad definition admits a wide range of actors that do not seem like computers, including consortia of crabs [234], slime molds [235], fluids [236], and even algorithms running inside other computers [237]. For all of these unconventional computers, as well as for the novel mechanical computing substrates discussed above, it is difficult to tell at which point they transition from "just physical materials" into computers. With continuous dynamical systems such as these, observers may choose different views from which the system appears to be acting more or less like a physical instantiation of a Turing Machine.

Even if an observed system seems to be behaving as if it is a Turing Machine, identifying the components of that machine, such as the tape or the read/write head, can be difficult. This is a common reason why it is often claimed that organisms are not machines/computers [67,238,239]. Consider an example from the authors' own recent work [72]. We found that motile multicellular assemblies can "build" other motile assemblies from loose cells. This looks very much like von Neumann machines: theoretical machines that can construct copies of themselves from materials in their environment. von Neumann initially proved the possibility of such machines by constructing, mathematically, Turing Machines that built copies of themselves by referring to and altering an internal tape. However, in the biological Turing Machines we observed, there seems to be no tape. If there is one, it is not likely localized in space and time.

This difficulty in identifying whether something is a computer, or at what point it becomes one, is further frustrated by the fact that biological and non-biological systems change over time:

even if one holds one view constant, the system, as it changes, may seem to act more or less like a computer. Finally, a polycomputing system, because it can provide different computational results to different observers simultaneously, may at the same time present as different computers --- better or worse ones, more general or more specialized ones --- to those observers. Such behavior would not only foil the question "Is that a computer?", but would even foil any attempts to determine the time at which a system becomes a computer, or begins to act more like a computer. Zooming out, it seems that as more advanced technology is created, and our understanding of biological systems progresses, attempts to attribute any singular cognitive self to a given system will become increasingly untenable. Instead, we will be forced, by our own engineering and science, to admit that many systems of interest house multiple selves, with more or less computational and agential potential, not just at different size scales, but also superimposed upon one another within any localized part of the system.

As Hoel points out about the ad-hoc status of claiming one single privileged Perspective within a system according to the Integrated Information Theory (IIT) account of consciousness [240]:

> "…There are so many viable scales of description, including computations, and all have *some* degree of integrated information. So, the exclusion postulate is necessary to get a definite singular consciousness. This ends up being the most controversial postulate within IIT, however." [94]

We hold that as our understanding of polycomputing biological and technological systems increases, it will eventually exclude the exclusion postulate from any attempt to explain human consciousness as some mental module operating within the brain.

## 7. Conclusion

Prior skeptical debates about whether biological systems are computers reflect both an outdated view of computation and a mistaken belief that there is a single, objective answer. Instead, we suggest a view in which computational models are not simply lenses through which diverse observers can all understand a given system in the same way, but indeed that several computational models can be true of a biological system at the same time. It is now seen that there is no one-to-one mapping between biological form and function: the high conservation of biological form and function across evolutionary instances implements a kind of multiple realizability. At the same time, biological components are massively overloaded toward polycomputing. Indeed, their competency, plasticity, and autonomy [2,241-243] may enable a kind of second-order polycomputing where various body components attempt to model each other's computational behavior (in effect serving as observers) and act based on their expected reward. Thus, modern computer engineering offers metaphors much more suited to understand and predict life than prior (linear, absolute) computational frameworks. Not only are biological systems a kind of computer (an extremely powerful one), but they are amazing *polycomputing* devices of a depth which has not yet been achieved by technology. In this sense, biological systems are indeed different than today's computers, although there is no reason why future efforts at deep, multiscale, highly plastic synthetic devices cannot take advantage of the principles of biologic polycomputing. A key implication of our view is that that blanket pronouncements about what living, or non-living, machines can do are worthless: we are guaranteed to be surprised by outcomes that can only be achieved by formulating and testing hypotheses. It is already clear that synthetic, evolved, and hybrid systems far outstrip our ability to predict limits on their adaptive behavior; abandoning absolutist categories and objective views of computation is a first step towards expanding our predictive capabilities.

At stake are numerous practical outcomes in addition to fundamental questions. For example, to make transformative advances in our ability to improve health in biomedical settings [244], we must be able to control multiple scales of biological organization which are heavily polycomputing - from cellular pathways to patient psychological state. It is essential to begin to develop computational frameworks to facilitate that kind of control. The ability to construct and model a kind of computational superposition, in which diverse observers (scientists, users, the agent itself, and its various components) have their own model of their dynamic environment and optimize their behavior accordingly, will also dovetail with and advance efforts in synthetic bioengineering, biorobotics, smart materials, and AI.

**Funding**

M.L. gratefully acknowledges the support of the Templeton World Charity Foundation (grant TWCF0606) and the John Templeton Foundation (grant 62212). J. B. gratefully acknowledges the support of the National Science Foundation (NAIRI award 2020247; DMREF award 2118988).

**Acknowledgements**

We thank Oded Rechavi and Aimer G. Diaz for useful pointers to relevant biological phenomena, and Julia Poirier and Susan Lewis for assistance with the manuscript.

**Author Contributions**

J.B. and M.L. contributed equally to this work.

**Conflict of Interest**

M.L. and J.B. are co-founders of Fauna Systems, an AI-biorobotics company; we declare no other competing interests.